\documentstyle[12pt,aps]{revtex}

\def\aap{{Astronomy and Astrophys.}}	
\def\jgr{{J.~Geophys.~Res.}}

\def\apj{{\it Astrophys. J.\ }}
\def\apjs{{\apj\ \it Suppl.\ }}		
\def\apjl{{\apj\ \it Lett.\ }}

\def\app{{\it Astropart. Phys.\ }}		
\def\pre{{\it Phys.~Rev.~E\ }}

\def\prl{{\it Phys.~Rev.~Lett.\ }}	
\def\jgr{{\it J.~Geophys.~Res.\ }}

\def\physrep{{\it Phys.~Rep.\ }}

\def\pasp{{\it Publ. Astron. Soc. Pac.\ }}

\def\la{\lesssim}

\def\etal{{\it et al.,\ }}
\def\eg{{e.g.,\ }}
\def\ie{{i.e.,\ }}

\begin{document}

\title{Modern theory of Fermi acceleration: a new challenge to plasma physics}

\author{M.A. Malkov and P.H. Diamond}

\address{University of California at San Diego, 
9500 Gilman Dr, La Jolla, CA 92093-0319, USA}
\maketitle

\begin{abstract}
One of the main features of astrophysical shocks is their ability to
accelerate particles to extremely high energies. The leading
acceleration mechanism, the diffusive shock acceleration is
reviewed. It is demonstrated that its efficiency critically depends on
the injection of thermal plasma into acceleration which takes place at
the subshock of the collisionless shock structure that, in turn, can
be significantly smoothed by energetic particles. Furthermore, their
inhomogeneous distribution provides free energy for MHD turbulence
regulating the subshock strength and injection rate. Moreover, the MHD
turbulence confines particles to the shock front controlling their
maximum energy and bootstrapping acceleration. Therefore, the study of
the MHD turbulence in a compressive plasma flow near a shock is a key
to understanding of the entire process. The calculation of the
injection rate became part of the collisionless shock theory.  It is
argued that the further progress in diffusive shock acceleration
theory is impossible without a significant advance in these two areas
of plasma physics.
\end{abstract}
\pagebreak


\section{Introduction\label{intr-mot} }

Over the last few years new observations and missions, \eg Energetic
Gamma-ray Experiment Telescope (EGRET) \cite{egr99}, \emph{Chandra}
\cite{hug00}, TeV- astronomy \cite{ah99}, revolutionized the
measurement of radiation from a variety of objects in the Universe. In
most cases the \textit{primary} source of the radiation is believed to
be accelerated charged particles, often of remarkably high energies,
such as \( 10^{20}eV \) or even higher \cite{bla00}, usually referred
to as the ultra high energy cosmic rays (UHECR). The accelerated
particles themselves (we will also use the term cosmic rays, CRs) are
in many cases generated by shock waves (shocks). Note, that the latter
are the major events where the huge energy of stars, Supernovae's (SN)
or Black holes is released in bulk gas motions and ultimately
dissipated. The most successful particle acceleration mechanism is,
perhaps the diffusive shock acceleration (DSA) \cite{dru83,be87},
which is a variant of the original Fermi idea (1949), also known as
the First order Fermi acceleration process. According to this
mechanism particles gain energy by bouncing off hydromagnetic
disturbances frozen into the converging upstream and downstream flow
regions near a shock. Clearly, the understanding of this mechanism is
critical for radiation models since the primary particle spectrum is
one of their most important ``input'' characteristics. The success of
this mechanism has been due to the following appealing features

\begin{enumerate}
\item it reproduces the power-law energy spectra with an 
index remarkably similar to that inferred
from the observations of the galactic cosmic rays (CRs)
\cite{dru83,be87}.
\item convincing, direct observational evidence of its 
operation at interplanetary shocks and the
earth bow shock \cite{fisk,lee82,kenetal84} 
\item the absence of any obvious intrinsic limitation 
to the maximum particle energy achievable by
this process 
\end{enumerate}
Often, however, this mechanism is applied in its simplified version,
the so called test particle (TP) or linear approximation (that
neglects the backreaction of accelerated particles on the shock
structure and produces a simple \( E^{-2} \) particle energy
distribution). Although more realistic, some nonlinear theories do
exist, they suffer from parameterization of particle transport
coefficients such as the pitch-angle scattering, spatial diffusivity
and plasma heating.

Also, it should be emphasized that the statement (2) above is only a
prima facie evidence for the responsibility of this mechanism for
galactic CRs, alluded to in (1). Despite similarity between physical
parameters \cite{be87}, the life-times and extensions of shocks in
these two environments are vastly different (up to a factor \( \sim
10^{10} \)). This requires different approximations for these
shocks. Namely, the treatment of large astrophysical shocks such as
Supernova remnant (SNR) shells, shocks in the lobes of radio galaxies
or even larger shocks in clusters of galaxies \emph{must} include the
back reaction of accelerated particles on the shock structure, and
thus be \textit{intrinsically nonlinear}. This, in turn, necessitates
the selfconsistent treatment of the above mentioned anomalous
transport phenomena.

From the perspective of the widely accepted SNR shock hypothesis of
the origin of galactic CRs, perhaps, the most critical impact of this
treatment would be on the statement (3) above.  It implies that the
maximum energy is limited only by the time available for acceleration
and by the size of accelerating object (roughly not to be exceeded by
the diffusive escape length for particles, which grows with
energy). This is becoming a ``hot'' issue in view of the lack of
evidence for TeV-protons in SNR shocks (despite the signature of
GeV-protons, EGRET), which might indicate that the particle spectrum
either cuts off somewhere between GeV and TeV energies (\ie in a
currently uncovered by any instrument energy range) or probably
significantly steepens there. We must await observations from
Gamma-ray Large Area Space Telescope (GLAST) and the next generation
of the imaging atmospheric Cherenkov telescopes that are currently
being built \cite{ah99} to see where and how the spectrum disappears
or changes. To understand how this \emph{may} happen, one needs to
consider all the crucial requirements for this mechanism, which are

\begin{itemize}
\item the ability of some fraction of thermal particles downstream 
to make the first step to acceleration,
\ie to return upstream (injection problem) 
\item good confinement of accelerated particles near the shock 
front (to continue energy gain) 
\item the ability of the shock to withstand the pressure of 
accelerated particles (the shock smoothing,
or shock robustness problem) 
\end{itemize}
These requirements are also associated with the main difficulties of
the theory and are, in fact, strongly related. We consider them
briefly, in order.

\subsection{Injection problem}

Injection is the process of initial particle energization. Within the
widely accepted ``thermal leakage'' scenario of injection, the way
protons enter the shock acceleration process is physically the same as
when they are accelerated afterwards (see, \eg
Ref.\cite{quest}). After thermalization downstream, certain fraction
of particles will catch up with the shock. Their further leakage
upstream generates Alfven waves via the cyclotron resonance \( kp\mu
\approx eB/c \) where
\( \mu  \) is the cosine of the proton pitch angle, and \( p \) is 
the particle momentum.  These waves do two things. First, they
self-regulate the thermal particle leakage (injection) by trapping
particles downstream when their leakage becomes too strong and
therefore the wave amplitude too large as well \cite{m98}. Second, the
large amplitude waves scatter already accelerating particles in pitch
angle, thus ensuring their diffusive confinement near the shock.

According to the above resonance condition, waves excited by protons
(with \( k\rho _{p}\sim 1) \) are too long to scatter electrons, so
that they interact with them adiabatically (\( k\rho _{e}\ll 1 \)).
Therefore, electrons need a separate injection scenario. One
suggestion is that they might be injected via scattering on
self-generated whistler waves \cite{lev92}. According to another
mechanism suggested in Ref.\cite{gmv95}, electrons are extracted
directly from the thermal pool near the shock front via their
interaction with proton-generated lower-hybrid waves and electrostatic
barrier formed there due to this interaction.

It should be noted that when the both species become relativistic and,
if the synchrotron cooling time is longer than the acceleration time,
the electron spectrum should be identical to that of the
protons. However, it is electrons for which we have convincing direct
evidence of Fermi acceleration in SNR shells. There are well
documented measurements of nonthermal electron emission in radio, \( x
\) rays and possibly also in \( \gamma \) rays, presented \eg in
Refs. \cite{green91},
\cite{koy95} and \cite{tanim98}, respectively. Although electrons 
play perhaps dynamically unimportant role in the shock structure, they
should trace the dynamically important proton spectra.

\subsection{Particle Confinement near the shock \label{conf} }

As in the injection phase, further particle confinement near the shock
is supported by self-generated Alfven wave turbulence, since the
energetic particle distribution ahead of the shock is ion-cyclotron
unstable (see, \eg Ref.\cite{lee82}). Due to rapid pitch-angle
scattering on these waves, particles cross the shock repeatedly, thus
gaining energy. Clearly, the scattering frequency (turbulence level)
sets up the acceleration rate. Note that the existing background
magneto-hydrodynamic (MHD) turbulence in the interstellar medium (ISM)
would support only very slow acceleration.  A long-standing problem is
that the wave generation process is likely to be so robust that wave
amplitude may far exceed the level admissible by the quasi-linear
theory. It is usually expected that the turbulence saturates at a
level \( \delta B\sim B_{0} \). Thus, particle scattering must occur
via strongly nonlinear wave-wave and wave-particle interactions so
that the conventional quasi-linear description should be replaced by a
non-perturbative approach.

MHD simulations significantly advanced recently (see, eg \cite{bisk}
and references therein) but they would not suffice alone to
self-consistently describe the wave generation by turbulently confined
particles. The main difficulty is, as it will be seen in the sequel,
the enormous extension of particle and wave spectra. It seems
desirable to combine simulations in restricted but critical parts of
the phase space (as the short-wave, low-energy part that controlls the
injection and the long-wave, high energy part where particle losses
occur) with an analytical approach in extended but tractable parts of
the phase space where particle and wave spectra and even the flow
profile exhibit relatively simple, scale invariant behaviour
\cite{md00}.

It should be evident that a purely numerical approach would encounter
serious difficulties if applied to the shock acceleration of
UHECRs. Indeed, even if we had numerical solution of the problem of CR
acceleration in SNR shocks, \ie up to energies \( \sim 10^{15} \) eV,
we would need to extend its dynamical range by a factor of \( \sim
10^{5} \). Since the integration domain in configuartion space is
typically proportional to the maximum energy, this would mean
\( \sim 10^{10} \) times larger 
phase space. It seems natural to use this large size of the
phase space instead of letting computer to process decade after decade
of it with basically the same physics. The analytic approach that we
introduce in Sec.\ref{kin-th} below utilizes the corresponding small
parameters and therefore appears particularly suitable to the problem
of acceleration of UHECRs.

\subsection{Shock robustness \label{robust} }

Due to the secular growth of the pressure of accelerated particles,
the linear solution becomes invalid for shocks of sufficient life
time, scale \( L \) and magnetic field \( B \) (the most critical
parameter is \( BL \) since the maximum particle energy scales as \(
E_{max}\sim (u/c)eBL \), provided that enough time is available for
acceleration, as discussed, \eg in Ref.\cite{be87}).  The main
nonlinear effect is due to the back reaction of energetic particles on
the flow which reduces the velocity jump at the flow discontinuity
(subshock) through the deceleration and heating of the plasma in front
of it (\ie in the so called CR precursor), Fig.\ref{fig:setup}.  The
total shock compression, however, increases due to the decrease in the
adiabatic index
\( \gamma  \) caused by the presence of relativistic particles and, 
even more importantly,
by their escape from the system (akin to radiative shocks). At the
same time the maintenance of a finite subshock within the global shock
structure (subshock itself, plus CR precursor) is critical to the
injection process and thus for acceleration in general. A mathematical
limitation of the TP approximation can be obtained in terms of
particle energy and injection rate \( \nu \), namely, \( \nu
\sqrt{E_{max}}\la 1 \) \cite{m97a}, which states that the pressure of
accelerated particles must remain smaller than the shock ram
pressure. It is also clear from this condition that the above three
issues are strongly coupled. Injection (\( \nu \)) and particle
confinement (which sets \( E_{max} \)) determine the shock structure
and are, in turn, regulated by it.  The acceleration time scale \(
\tau _{acc} \) (or particle diffusivity \( \kappa \)) and the
precursor turbulent heating rate are also implicitly involved in this
``feedback loop''.  There are indeed too many shock variables to
obtain reliable prediction by simply scanning the parameter
space. Therefore, they need to be calculated self-consistently before
the particle spectra can be calculated and compared with the
observations.

\section{Nonlinear theories \label{nl-th} }

A typical nonlinearly accelerating shock is illustrated in
Fig.\ref{fig:setup}. The most ``visible'' nonlinear effects are: (1)
the deceleration of the flow upstream (see the flow profile on the top
of Fig. \ref{fig:setup}) by the pressure of accelerated particles, (2)
subshock reduction, as a consequence (its strength \( u_{0}/u_{2} \)
may become small compared to the total compression \( u_{1}/u_{2} \)
(3) bending of the B-field due to the compression of its tangential
component because of the frozen in condition, \( uB_{t}=const \),
while the normal component is conserved due to \( {\rm
div}\mathbf{B}=0 \).

\subsection{Two-fluid model }

Initially, the back reaction of accelerated particles (CRs) on the
shock structure was studied within the two-fluid model (TFM). This
model treats CRs as a second fluid carrying the momentum and energy
across the shock, but not the mass. Complete solutions given in
Refs.\cite{dv81,alm82} indeed revealed a very strong back reaction of
the CRs onto the bulk plasma flow, leading to a bifurcation of the
simple linear solution into a strongly nonlinear (efficient) solution
with the acceleration efficiency approaching (in strong shocks) 100\%
due to the nonlinearly enhanced shock compression. This gives rise to
a formally diverging (\ie in reality strongly dependent on the maximum
energy) CR pressure which makes the particle losses at the highest
energies as important dynamically as injection, and strongly related
to it through their feedback on the subshock strength. Moreover, these
losses are controlled by the Alfven turbulence which, in turn, depends
on the CR distribution in the shock precursor and thus on the losses
themselves.  Unfortunately, the TFM, being a hydrodynamic theory,
cannot be closed in such a way that these essentially kinetic effects
are properly represented.

\subsection{Kinetic Theory \label{kin-th} }

The earlier kinetic theories also demonstrated that {\it upon
accumulating enough CR energy, strong shocks develop into the
nonlinear regime} The minimal kinetic theory
that captures bifurcations of the DSA can be formulated as follows.

We describe the distribution of high energy particles (CRs) by the so
called diffusion-convection equation (see, \eg
Refs.\cite{dru83,be87}). The gaseous discontinuity (the subshock) is
assumed to be located at \( x=0 \) and, for convenience, we flip the
\( x \)- coordinate in Fig.\ref{fig:setup}, so that the 
upstream side is \( x>0 \) half-space. 
Thus, the flow velocity
in the shock frame can be represented as \( V(x)=-u(x) \) where the
(positive) flow speed
\( u(x) \) jumps from \( u_{2}\equiv u(0-) \) downstream 
to \( u_{0}\equiv u(0+)>u_{2} \)
across the subshock and then gradually grows upstream up to \(
u_{1}\equiv u(+\infty )\geq u_{0} \).  In a steady state the equation
can be written as
\begin{equation}
\label{c:d}
u\frac{\partial f}{\partial x}+\kappa (p)\frac{\partial ^{2}f}
{\partial x^{2}}=\frac{1}{3}\frac{du}{dx}p\frac{\partial f}{\partial p},\protect 
\end{equation}
where \( f(x,p) \) is the isotropic (in the local fluid frame) part of
the particle distribution.  This is assumed to vanish far upstream (\(
f\rightarrow 0,\, x\rightarrow \infty \)), while the only bounded
solution downstream is obviously \( f(x,p)=f_{0}(p)\equiv f(0,p)
\). The most plausible assumption about the cosmic ray diffusivity \(
\kappa (p) \) is that of the Bohm type, \ie \( \kappa
(p)=Kp^{2}/\sqrt{1+p^{2}} \) (the particle momentum \( p \) is
normalized to \( mc \)). In other words \( \kappa \) scales as the
gyroradius, \( \kappa \sim r_{\rm g}(p) \).  The reference diffusivity
\( K \) depends on the \( \delta B/B \) level of the Alfvenic
turbulence that scatters the particles in pitch angle. The minimum
value for \( K \) would be \( K\sim mc^{3}/eB \) if \( \delta B\sim B
\). Note that the replacement of this plain parameterization by a
selfconsistent solution for the spectrum of turbulence driven by an
inhomogeneous distribution of accelerated particles is a challenge to
plasma physics. The existing quasi-linear approaches clearly fail in
the efficient acceleration regime due to unacceptable wave amplitudes.

The determination of \( u(x) \) in eq.(\ref{c:d}) requires three
further equations. The first one is the conservation of the momentum
flux in the smooth part of the shock transition (\( x>0 \),
\ie in the CR-precursor) 
\begin{equation}
\label{mom:c}
P_{\rm c}+\rho u^{2}=\rho _{1}u_{1}^{2},\quad x>0
\end{equation}
 where \( P_{\rm c} \) is the pressure of the high energy particles 
\begin{equation}
\label{P_{c}}
P_{\rm c}(x)=\frac{4\pi }{3}mc^{2}\int _{p_{0}}^{p_{1}}
\frac{p^{4}dp}{\sqrt{p^{2}+1}}f(p,x)
\end{equation}
It is assumed here that there are no particles with momenta \( p>p_{1}
\) (they leave the shock vicinity because there are no MHD waves with
sufficiently long wave length \( \lambda \), since the cyclotron
resonance requires \( p\sim \lambda \)). The momentum region \(
0<p<p_{0} \) cannot be described by equation (\ref{c:d}) and the
behavior of \( f(p) \) at \( p\sim p_{0} \) is described by the
injection parameters \( p_{0} \) and \( f(p_{0}) \) \cite{m97a}. The
plasma density \( \rho (x) \) can be eliminated from equation
(\ref{mom:c}) by using the continuity equation \( \rho u=\rho
_{1}u_{1} \). Finally, the subshock strength \( r_{\rm{s}} \) can
be expressed through the Mach number \( M \) at \( x=\infty \)
\begin{equation}
\label{c:r}
r_{\rm s}\equiv \frac{u_{0}}{u_{2}}=\frac{\gamma +1}
{\gamma -1+2R^{\gamma +1}M^{-2}}
\end{equation}
where the precursor compression \( R\equiv u_{1}/u_{0} \) and \(
\gamma \) is the adiabatic index of the plasma.

The system of equations (\ref{c:d},\ref{mom:c},\ref{c:r}) describes in
a self-consistent manner the particle spectrum and the flow structure,
although under parameterization of such critical quantities as \( \nu
\) and \( p_{1} \). Our poor knowledge of the maximum momentum
\( p_{1} \) is related to the prescribed form of 
particle diffusivity \( \kappa (p) \). 

It is useful to reduce this system to one integral equation
\cite{m97a}. A key dependent variable is an integral transform of the
flow profile \( u(x) \) with a kernel suggested by an asymptotic
solution of the system (\ref{c:d})-(\ref{mom:c}) which has the form
\[
f(x,p)=f_{0}(p)\exp \left[ -\frac{q}{3\kappa }\Psi \right] \]
 where 
\[
\Psi =\int _{0}^{x}u(x')dx'\]
 is the flow potential and the spectral index 
downstream \( q(p)=-d\ln f_{0}/d\ln p \). The
integral transform is as follows 
\begin{equation}
\label{U}
U(p)=\frac{1}{u_{1}}\int _{0-}^{\infty }\exp \left[
-\frac{q(p)}{3\kappa (p)}\Psi \right] du(\Psi )
\end{equation}
 and it is related to \( q(p) \) through the following formula 
\begin{equation}
\label{q(p)}
q(p)=\frac{d\ln U}{d\ln p}+\frac{3}{r_{\rm{s}}RU(p)}+3
\end{equation}
The physical meaning of the function \( U(p) \) is very simple. It
reflects the degree of shock modification. Namely, a function \(
U(p)+u_{2} \) is an effective flow velocity upstream as seen by a
particle with momentum \( p \) that diffusively escapes ahead of the
shock to a point \( x \) where the flow speed is \( u(x)=U(p)+u_{2}
\). Once \( U(p) \) is found, both the flow profile and the particle
distribution can be determined by inverting transform (\ref{U}) and
integrating equation (\ref{q(p)}). Now, using the linearity of
equation (\ref{mom:c}) (\( \rho u=const \)), we derive the integral
equation for \( U \) by applying the transformation (\ref{U}) to the
\( x- \) derivative of equation (\ref{mom:c}) \cite{m97a}. The result
reads
\begin{eqnarray}
U(t) & = & \frac{r_{\rm{s}}-1}{Rr_{\rm{s}}}+\frac{\nu
 }{Kp_{0}}\int _{t_{0}}^{t_{1}}dt'\left[ \frac{1}{\kappa
 (t')}+\frac{q(t')}{\kappa (t)q(t)}\right] ^{-1}\nonumber \\ & \times
 & \frac{U(t_{0})}{U(t')}\exp \left[ -\frac{3}{Rr_{\rm{s}}}\int
 _{t_{0}}^{t'}\frac{dt''}{U(t'')}\right] \label{int:eq}
\end{eqnarray}
 where \( t=\ln p \), \( t_{0,1}=\ln p_{0,1} \). Here the injection parameter 
\begin{equation}
\label{nu:def}
\nu =\frac{4\pi }{3}\frac{mc^{2}}{\rho _{1}u_{1}^{2}}p_{0}^{4}f_{0}(p_{0})
\end{equation}
 is related to \( R \) by means of the following equation 
\begin{eqnarray}
\nu  & = & Kp_{0}\left( 1-R^{-1}\right) \nonumber \\
 & \times & \left\{ \int _{t_{0}}^{t_{1}}\kappa
(t)dt\frac{U(t_{0})}{U(t)}\exp \left[ -\frac{3}{Rr_{\rm{s}}}\int
_{t_{0}}^{t}\frac{dt'}{U(t')}\right] \right\} ^{-1}\label{nu}
\end{eqnarray}
The equations (\ref{c:r},\ref{int:eq},\ref{nu}) form a closed system
that can be easily solved, \eg numerically. Asymptotic analytic
solutions are also available \cite{m97a}. The physical quantities
involved are: the far upstream Mach number \( M \) (which is given,
external parameter); internal parameters: \( M_{0} \) (the Mach number
at the subshock), the injection rate \( \nu \), the particle maximum
momentum \( p_{max}\equiv p_{1} \), and particle diffusivity
\( \kappa (p,x) \). These internal parameters must be determined 
self-consistently, but currently
are parameterized or calculated using some simplifying assumption. For
example, assuming that there is no turbulent heating in the shock
precursor (which is doubtful in such a turbulent environment), the
parameter \( M_{0}=M/R^{(\gamma +1)/2} \) {[}see eq.(\ref{c:r}){]}.

\section{Critical nature of acceleration process}
The presence of bifurcation in this acceleration process is best seen
in variables \( R,\nu \).  The quantity \( R-1 \) is a measure of
shock modification produced by CRs, in fact \(
(R-1)/R=P_{\rm{c}}(0)/\rho _{1}u_{1}^{2} \) {[}Eq.(\ref{mom:c}){]}
and may be regarded as an order parameter. The injection rate \( \nu
\) characterizes the CR density at the shock front and can be
tentatively treated as a control parameter. It is convenient to plot
the function \( \nu (R) \) instead of \( R(\nu ) \) {[}using equations
(\ref{nu}) and (\ref{c:r}){]}, Fig.\ref{fig:NuOfR}.

In fact, the injection rate \( \nu \) at the subshock should be
calculated given \( r_{\rm{s}}(R) \) with the self-consistent
determination of the flow compression \( R \) on the basis of the
\( R(\nu ) \) dependence obtained. However, in view of a very 
strong, even nonunique dependence
\( R(\nu ) \), this solution can be physically meaningful only 
in regimes far from criticality,
\ie when \( R\approx 1 \) (test particle regime) or \( R\gg 1 \) 
(efficient acceleration).
There are, however, self-regulating processes that should drive the
system \emph{towards critical region} (where \( R(\nu ) \) dependence
is very sharp) \cite{md00,mdv00}. First, if \( \nu \) is subcritical
it will inevitably become supercritical when \( p_{1} \) grows in
course of acceleration. Once it happened, however, the strong subshock
reduction (equation {[}\ref{c:r}{]}) will reduce \( \nu \) as well and
drive the system back to the critical regime, Fig.\ref{fig:bif}.

The maximum momentum \( p_{1} \) is subject to self-regulation as
well. Indeed, when \( R\gg 1 \), the generation and propagation of
Alfven waves is characterized by strong inclination of the
characteristics of wave transport equation towards larger wavenumbers
\( k \) on the \( k-x \) plane due to wave compression. Thus,
considering particles with \( p\la p_{1} \) inside the precursor, one
sees that they are in resonance with waves that must have been excited
by particles with \( p>p_{1} \) further upstream but, there are no
particles with \( p>p_{1} \). Therefore, the required waves can be
excited only locally by the same particles with \( p\la p_{1} \) which
substantially diminishes the amplitude of waves that are in resonance
with particles from the interval \( p_{1}/R<p<p_{1} \). (The left
inequality arises from the resonance condition
\( kcp\approx eB/mc \) and the frequency conservation along the 
characteristics \( ku(x)=const \)).
This will worsen the confinement of these particles to the shock
front. The quantitative study of this process is another challenge to
the theory of plasma turbulence. What can be inferred from
Fig.\ref{fig:NuOfR} now, is that the decrease of \( p_{1} \)
straightens out and rise the curve \( \nu (R) \), so that it returns
to the monotonic behaviour. However once the actual injection becomes
subcritical (and thus \( R\rightarrow 1 \)) then \( p_{1} \) will grow
again restoring the two extrema on the curve \( \nu (R) \).

Also the turbulent precursor heating straightens up the bifurcation
diagram and returns it to the critical state. We illustrate this in
Fig.\ref{fig:heat} for different heating efficiencies
\( \alpha  \), introduced phenomenologically through the decrease 
in the flow Mach number
across the CR precursor

\begin{equation}
\label{nad:h}
M_0^{-2}=M^{-2}R^{8/3}+\frac{\alpha }{3}(R-1)p_{max}
\end{equation}
Here the first term is due to the familiar adiabatic compression while
the second describes the turbulent heating, assumed to be proportional
to the CR pressure contrast \( \propto R-1 \) (source of the free
energy for turbulence) as well as to the precursor length \( \propto
p_{max} \).  The plots in Figs. \ref{fig:NuOfR},\ref{fig:heat}
demonstrate that parameterization approaches in which the internal
parameters (like \( \alpha \) and \( p_{max} \) in many simulations)
are treated as given may be useful only when (and if) these parameters
are well outside the critical region which is marked by very sharp or
even nonunique parameter dependence. For example, as seen from
Fig.\ref{fig:heat}, if \( \alpha \) (which is essentially unknown)
lies within the interval \( 10^{-8}-10^{-7} \), then there are no
means to calculate the flow structure (\( R \)), and thus the particle
spectrum with a reasonable accuracy. To treat the injection rate in
the critical region as a ``control'' parameter is equally useless,
Figs. \ref{fig:NuOfR},\ref{fig:heat}.

It appears to be more productive to \emph{assume} a \emph{marginal
state} in which the maximum and the minimum of \( \nu (R) \) merge, at
least in a sense of an averaged (in time and space)
process.

This
means that \( \nu '(R)=\nu ''(R)=0 \) at some \( R=R_{\rm {c}}
\). These two equations and the dependence \( \nu (R) \) itself, not
only determine \( R_{\rm c} \) and \( \nu _{\rm c}\equiv \nu (R_{\rm
c}) \) but also provide an additional relation that involve other
parameters of the problem which clearly enter the function \( \nu (R)
\). These are the Mach number \( M \), the heating rate \( \alpha \)
and the maximum momentum \( p_{\rm {max}}\equiv p_{1} \). For example,
given \( M \) and \( \alpha \) we can easily calculate \( p_{\rm max}
\) \cite{mdv00}.  These results show, that at least in shocks with
high Mach numbers and no significant turbulent heating, the nonlinear
effects may significantly limit the maximum energy achievable by this
acceleration mechanism. Before any relevance of this result to, \eg the
lack of observational evidence of TeV protons in SNR can be seriously
debated, a number of difficult issues must be addressed which we
discuss briefly in the next section.

\section{Discussion}

Instead of relying on parameterization, and guessing about the
possible values of relevant parameters it was suggested that, due to
the self-regulation, the system should evolve precisely into the
critical point within given parameter space \cite{mdv00}. A
comprehensive evaluation of this suggestion and the study of
self-regulation mechanisms poses a serious challenge to plasma
physics. Note, however, that the critical self-organization approach
was proven very useful in describing transport phenomena in laboratory
plasmas, when the turbulence is generated by transport driving
gradients, \ie the pressure or density gradients, \eg \cite{diam94}.
The most important current issue is the construction of adequate links
between the internal parameters (and so the Alfvenic turbulence), and
their dependence upon the external parameters.  This is an
interdisciplinary problem in high energy astrophysics and nonlinear
plasma physics.  It combines hydrodynamics, particle kinetics with a
strong emphasis on the theory of dynamical chaos and ergodicity
(injection), Alfvenic turbulence, collisionless nonlinear plasma
phenomena and, in particular, the theory of collisionless shocks.

Currently not all the aspects of this problem are well understood. We
believe that we understand how the acceleration operates in the test
particle (linear) regime, and have an analytic description of
nonlinearly modified shocks including particle spectra, although under
a prescribed Bohm-type particle diffusion. Injection can be calculated
currently only for a relatively strong subshock, while we need it for
varying shock parameters of dynamically evolving CR shocks. Turbulence
dynamics, associated plasma heating and the turbulent transport of CRs
are understood to even lesser extent. A systematic determination of
parameters involved should allow the self-consistent calculation of
particle spectra.

Specifically, this requires detailed understanding of several physical
phenomena in a shock environment, which include:

\begin{enumerate}
\item generation of waves by energetic protons, wave transport and spectral evolution
\item turbulent transport of protons, in space and momentum, injection
\item electron injection triggered by proton generated turbulence
\item heating by Alfvenic and (magneto-) acoustic turbulence, 
which controls the global shock structure
\end{enumerate}
The resolution of these issues will lead to the further progress in
our understanding of particle acceleration and associated emission.

\section*{Acknowledgments}

Some of the presented results have been obtained in collaboration with
L. Drury and H. V\"olk.  We also acknowledge helpful discussions with
F. Aharonian and T. Jones.

This work was supported by U.S. DOE under Grant No. FG03-88ER53275.

\pagebreak

\pagebreak
\begin{figure}

\caption{Schematic representation of nonlinearly accelerating shock. 
Flow profile with a gradual deceleration
upstream is also shown at the top.\label{fig:setup}}
\end{figure}

\vspace{0.3cm}
{
\begin{figure}

\caption{Response of shock structure (bifurcation diagram) to the 
injection of thermal particles at
the rate \protect\( \nu \protect \). The strength of the response is
characterized by the pre-compression of the flow in the CR shock
precursor \protect\( R=u_{1}/u_{0}\protect \) (see Fig.1). The flow
Mach number \protect\( M=150\protect \); different curves correspond
to different values of maximum momentum denoted here as \protect\(
p_{1}\protect \)and normalized to \protect\( mc\protect \). For each
given \protect\( \nu \protect \) and \protect\( p_{max}\protect \) one
or three (arrows) solutions exist.\label{fig:NuOfR} }
\end{figure}
\begin{figure}

\caption{Bifurcation diagram corresponding to the set of response 
curves shown in Fig. \ref{fig:NuOfR}.  Since \protect\( \nu \protect
\) and \protect\( p_{max}\protect \) are in reality dynamic rather
than control parameters, the response curve moves towards the
bifurcation curve drawn with the heavy line. The resulting state of
the system corresponds to the critical (inflection) point on this
curve, which can be described as a ``self-organized critical'' (SOC)
state. \label{fig:bif} }
\end{figure}
\par}
\vspace{0.3cm}

\begin{figure}

\caption{Bifurcation diagrams as in Fig. 2 but for different heating 
rates \protect\( \alpha \protect \) (see text), for Mach number
\protect\( M=150\protect \) and the cut-off momentum \protect\(
p_{max}=10^{5}\protect \).
\label{fig:heat}}
\end{figure}

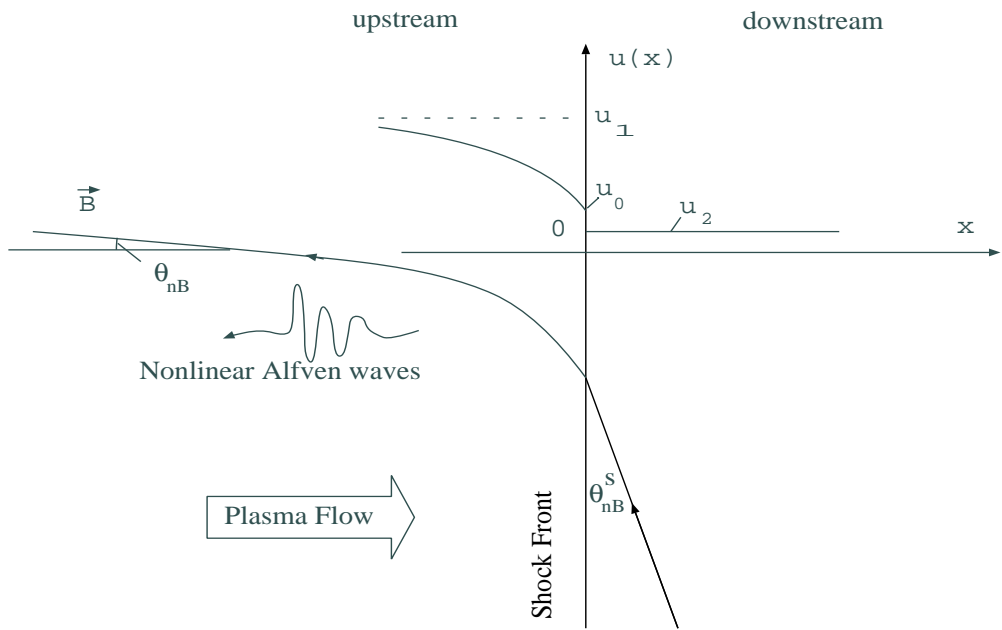
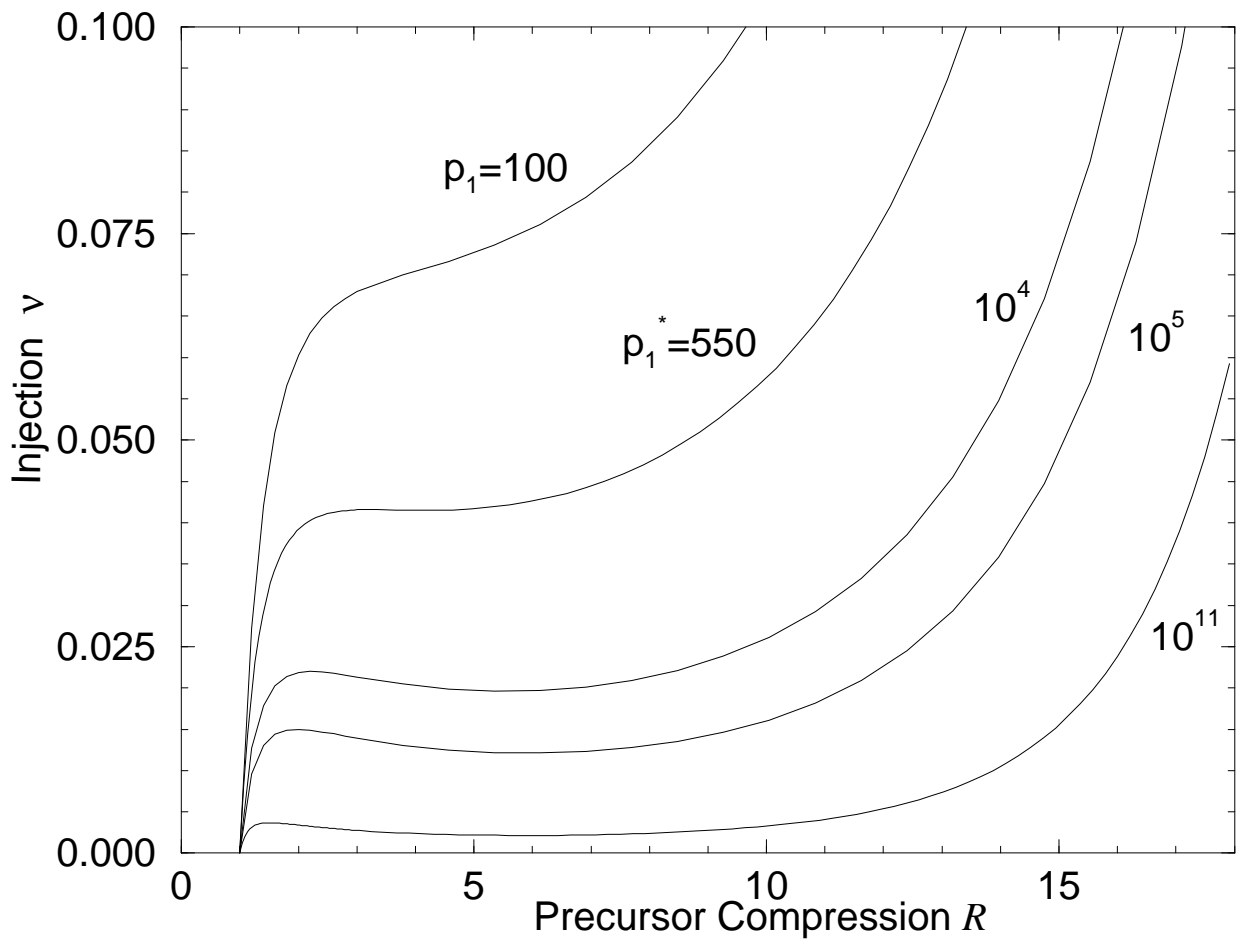
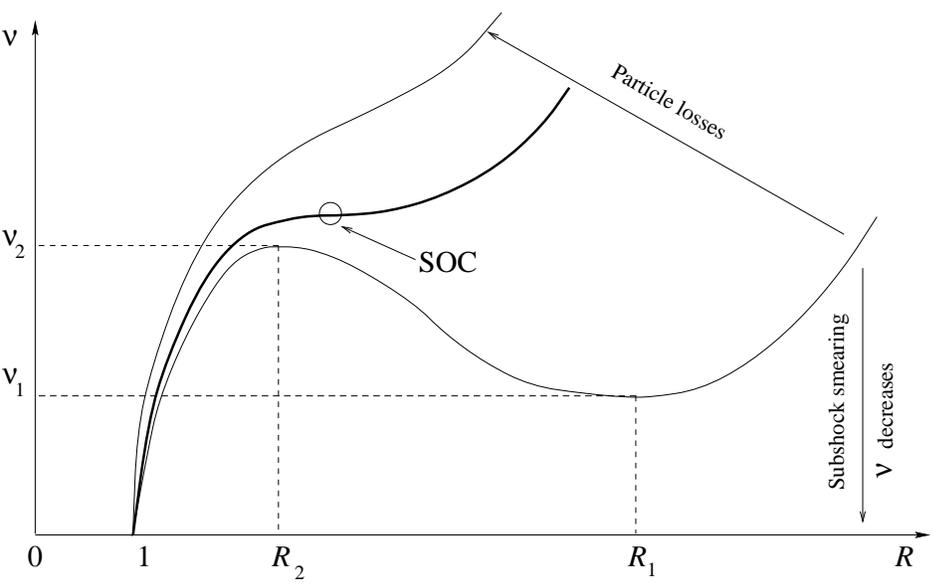
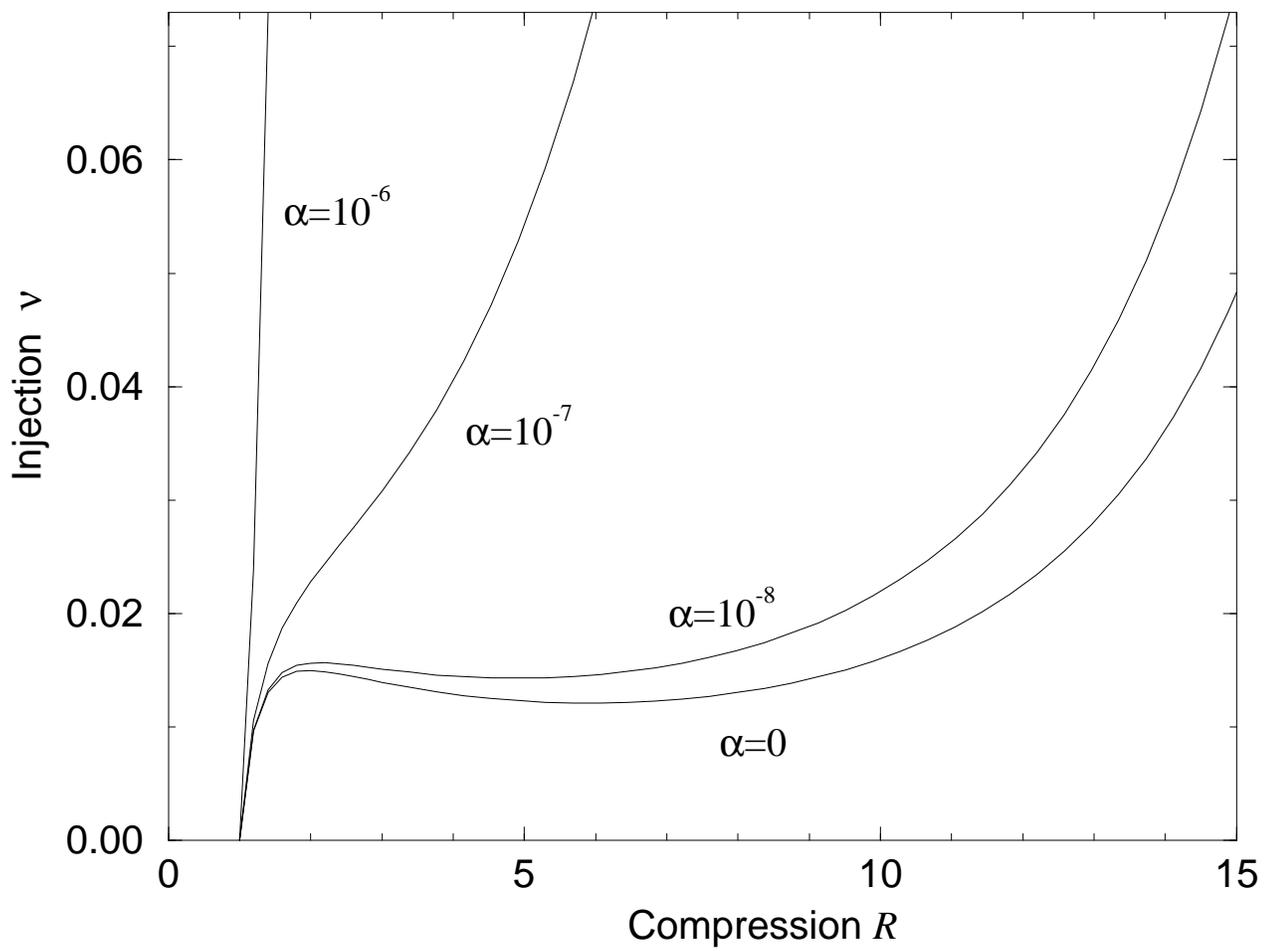
\end{document}